\documentclass[nofootinbib,notitlepage,superscriptaddress,10pt,aps,pra,twocolumn]{revtex4-1}

\usepackage[utf8x]{inputenc}
\usepackage{amsmath}
\usepackage{amssymb}
\usepackage{bbm}
\usepackage{graphicx}
\usepackage{mathptm}
\usepackage{color}

\begin{document}

\title{Possible relevance of quantum spacetime for neutrino-telescope data analyses}

\author{Giovanni Amelino-Camelia$^a$, Dafne Guetta$^{b,c}$ and Tsvi Piran$^d$\\
{\small{$^a$\it{Dipartimento di Fisica, Sapienza Universit\`a di Roma and INFN, Sez.~Roma1, P.le A. Moro 2, 00185 Roma, EU}}}\\
{\small{$^b$\it{Department of Physics, ORT Braude, Snunit 51 St. P.O.Box 78, Karmiel  21982, Israel}}}\\
{\small{$^c$\it{Osservatorio Astronomico di Roma, via Frascati 33, I00040 Monteporzio  Catone, Italy}}}\\
{\small{$^d$\it{The Racah Institute for Physics, The Hebrew University of Jerusalem, Jerusalem, 91904, Israel }}}}

\begin{abstract}
\noindent One of the primary goals of neutrino telescopes, such as IceCube, is the discovery of neutrinos
emitted by gamma-ray bursts (GRBs). Another source of interest in the results obtained by these telescopes
is their possible use for tests of the applicability of
Einstein's Special Relativity to neutrinos, particularly with respect to modifications
that lead to Lorentz invariance violation that have been
conjectured by  some models of quantum space-time.
We examine here the fascinating scenario in which these two aspects of neutrino-telescope physics
require a combined analysis. We discuss how
neutrinos that one would not associate to a GRB, when assuming a classical spacetime picture,
may well be GRB neutrinos if  the possibility that Lorentz invariance is broken at very high energies
is taken into account. As an illustrative example we
examine three IceCube  high energy neutrinos that arrived hours before GRBs (but from the same direction) and we
 find that the available,  IceCube data, while inconclusive,
 {is compatible with} a  scenario in which one or two of these neutrinos were
 GRB neutrinos and their earlier arrival reflects Lorentz invariance violation.
 We outline how future analyses of neutrino data should be done in order to systematically test this possibility.
\end{abstract}

\maketitle

\baselineskip11pt plus .5pt minus .5pt

Prominent on the agenda of the current generation of neutrino telescopes
is the search for neutrinos emitted in the same
gigantic explosion responsible for Gamma ray bursts (GRBs). This would surely mark a transition
to a rich new phase in understanding these most fascinating phenomena, adding
 the insight provided by the new neutrino window to
 the information available from electromagnetic observations.
The prediction of a neutrino emission associated with GRBs
is  generic within the  most widely accepted phenomenological interpretation of these
explosions,
given in terms of the so-called fireball model~\cite{fireball}.
But different variants of the model predict a different rate of neutrino production at
the GRB source. According to the fireball picture the energy
carried by the hadrons in a relativistic expanding wind is
dissipated through internal shocks between different parts of plasma.
These shocks reconvert a substantial part of the kinetic energy to
internal energy, which is then radiated as synchrotron and
inverse-Compton radiation of shock-accelerated electrons. When the
fireball has swept enough material it collides with its surrounding
medium giving rise to reverse and forward shocks, and the latter would then be
responsible for so-called afterglow emission\cite{meszaAFTERGLOW}.
Within this picture GRBs should produce   neutrinos
with energy of  $\sim 100$ TeV through the interaction of high
energy protons  with radiation, at the
same region where GRB-photons are produced\cite{dafneRATE}.
Neutrinos may be produced also in other stages
of fireball evolution and in particular within the afterglow or while a relativistic jet is still propagating
within the stellar envelope~\cite{meszawax}.

{Recently IceCube reported \cite{icecubenature} no detection of any  GRB-associated neutrino in a data set taken from April 2008 to May 2010.
 This comes in conflict with earlier predictions \cite{waxbig,meszabig,dafnebig,otherbig}, that predicted about 10 GRB neutrinos during this period. Those earlier estimates were largely calibrated assuming that Ultra-High Energy Cosmic Rays (UHECRs) are produced by GRBs. The IceCube results then appear to rule out GRBs as  the main sources of UHECRs  or that the efficiency of neutrino production is much lower than had been estimated
\cite{small1,small2,small3}. }

Since this issue ties in some of the most interesting
and hotly debated aspects of high-energy astrophysics,
it is interesting to explore alternatives to the conclusion suggested
by this analysis \cite{icecubenature}.
Of interest for the study we are here reporting is the fact that these assessments of the outcome of IceCube's GRB-neutrino
searches are based on the expectation that such neutrinos should be detected in  a  temporal coincidence with
the associated  $\gamma$-rays or the early afterglow.
We came to wonder how much of a difference it would make if the same data
were analyzed from the perspective of a Lorentz-invariance-violation scenario for propagation of GRB neutrinos first proposed by Jacob and Piran
\cite{jacobpiran} and more recently highlighted in an overall assessment of quantum-spacetime phenomenology
by Amelino-Camelia and Smolin \cite{gacsmolin}.
This scenario is inspired by research on violations of Lorentz symmetry seeded
in quantum properties of spacetime
and suggests that GRB neutrinos with energies of a few TeVs and above could be detected systematically much in advance
or much after the accompanying electromagnetic signal.

We shall summarize here the key ingredients of this picture momentarily,
but first let us observe that in the first IceCube data set, the IC40 data set,
the two most significant candidate GRB neutrinos were  both sizably in advance of the
trigger of the accompanying electromagnetic signal:
these were \cite{icecubetesi}
a 1.3 TeV neutrino {  1.95$^o$ off GRB090417B, with localization uncertainty of 1.61$^o$, and detection time  2249 seconds before the  trigger of GRB090417B, and
a 3.3 TeV neutrino  6.11$^o$ off GRB090219, with a localization uncertainty of 6.12$^o$, and detection time  3594 seconds before the GRB090219 trigger.
Neither of these two candidate GRB neutrinos could carry much significance, since they may well both be
just a (not-unlikely\cite{icecubetesi}) chance fluctuation of the background noise constituted by atmospheric neutrinos,
but we  nonetheless take notice of them.

For the other IceCube dataset \cite{icecubenature,icecubetesi}, the  IC59 data set,
two other events were highlighted by the IceCube collaboration, a 35TeV neutrino within 30 seconds of
GRB091026A, 4.5 degrees off-source, with a localization uncertainty of 10.5$^o$, and a 109 TeV neutrino, within 0.2$^o$  of GRB091230A, with a localization uncertainty of 0.2$^o$, and detected some 14 hours before the GRB091230A trigger.
While both these events were labelled as very likely cosmic-ray events
rather than GRB neutrinos \cite{icecubenature}, in a more detailed account \cite{icecubetesi}
it is observed that the 35 TeV event was very clearly a cosmic ray since it triggered the IceTop surface array, whereas
 for the 109 TeV event there was only one  IceTop-tank trigger in time coincidence. This single IceTop-tank trigger
may suggest it was part of a cosmic-ray air shower but \cite{halzen} could also be a background in the tank's photomultiplier. Indeed  the  109 TeV event has been described \cite{icecubeCERN}
as the most significant GRB-neutrino candidate so far reported by IceCube,
even combining both the IC40 and the IC59 data sets.  Following these remarks we exclude the 35 TeV event but
we include the 109 TeV event in our analysis. }

The fact that all  3 GRB-neutrino candidates
were detected sizably in advance of the triggers of the GRBs they could be associated with
is not particularly significant from the standard perspective of this sort of analysis,
and actually obstructs any such attempt to view them as  GRB neutrinos:
no current GRB model suggests that neutrinos could be emitted thousands of seconds before a GRB.
But a collection of GRB-neutrino candidates all sizably in advance of (or all with a sizable delay with respect to)
corresponding GRB triggers is just what one was expecting on the basis of the quantum-spacetime-inspired
Lorentz invariance violation scenario  \cite{jacobpiran,gacsmolin},
and this may invite further analysis.

This scenario for the discovery of GRB neutrinos \cite{jacobpiran,gacsmolin} was based on results
for models of spacetime quantization suggesting
that (see, {\it e.g.}, \cite{grbgac,gampul,urrutia,gacmaj,myePRL})  it is possible for the quantum properties of spacetime to introduce small violations of the special relativistic properties of classical spacetime.
A key consequence of this picture would be that
 the time needed for a ultrarelativistic particle\footnote{Of course the only regime of particle propagation
 that is relevant for this manuscript is the ultrarelativistic regime, since photons have no mass and
 for the neutrinos we are contemplating (energy of
 a few TeVs and above) the mass is completely negligible.}
to travel from a given source to a given
detector is  $t=t_0+t_{LIV}$. { Here $t_0$ is the time that would be predicted
in classical space-time,
while $t_{LIV}$ is the contribution to the travel time due to quantum properties of spacetime.
 For energies much smaller than, $M_{LIV}$,  the  scale  of onset of these quantum-spacetime effects,
 one expects that at lowest order} $t_{LIV}$ is given by  \cite{jacobpiranDELAY}:
\begin{equation}
t_{LIV} = - s_\pm \, \frac{E}{M_{LIV}} \frac{D(z)}{c}  ,
\label{main}
\end{equation}
where
\begin{equation}
D(z) = \int_0^z d\zeta \frac{(1+\zeta)}{H_0\sqrt{\Omega_\Lambda + (1+\zeta)^3 \Omega_m}}  .
\nonumber
\end{equation}
Here the information cosmology gives us on spacetime curvature is coded
in the denominator for the integrand
in $D(z)$, with $z$ being the redshift and $\Omega_\Lambda$, $H_0$ and $\Omega_0$ denoting, as usual,
respectively the cosmological constant, the Hubble parameter and the matter fraction.
The ``sign parameter" $s_\pm$, with allowed values of $1$ or $-1$, as well as the scale $M_{LIV}$  would have to be determined
experimentally. The label ``LIV'' stands for Lorent-invariance Violation, since the aspects of special relativity
 here at stake are indeed those connected to Lorentz invariance \cite{grbgac,gampul,urrutia,gacmaj,myePRL,jprd}
 (and there is interest in this class of effects from the intrinsic Lorentz-invariance test theory perspective \cite{mattiLRR},
 with or without spacetime quantization). We must stress however that most theorists
favor naturalness arguments suggesting that $M_{LIV}$ should take a value
that is
rather close to
the ``Planck scale"
$M_P =\sqrt{{\hbar c^5}/{G_N}} \simeq 1.22 \cdot 10^{16}TeV~.$

 The picture of quantum-spacetime effects
summarized in (\ref{main}) does not apply to all quantum-spacetime models. One can evisage quantum-spacetime pictures
that do not violate Lorentz symmetry at all, and even among the most studied quantum-spacetime pictures
that do violate Lorentz symmetry one also finds
variants producing (see, {\it e.g.}, \cite{gacsmolin,grbgac,mattiLRR})  features analogous to (\ref{main}) but with
the ratio $E/M_{LIV}$ replaced by its square, $(E/M_{LIV})^2$, in which case the effects
would be much weaker and practically undetectable at present.
We  focus here on the most studied Lorentz-invariance-violating scenario, the one centered on (\ref{main}).

It is important that  some quantum-spacetime models allow for laws roughly of the type (\ref{main})
to apply differently to photons and neutrinos.
An attractive hypothesis \cite{myePRL,mattiLRR}
is that the quantum-spacetime effects should still be accommodated within the
formalism of effective quantum field theory, where effects of the type
shown in  (\ref{main}) would take the shape of dimension-5 operators added to the Lagrangian density and contributing to the particle's propagator. Within
 that effective-field-theory setup one can formulate exactly (\ref{main}) for neutrinos, but not for photons (though a variant of (\ref{main}) with
an added polarization dependence is allowed for photons).
And even among quantum-spacetime models that do not fully comply with the demands of a  description within the effective-field-theory framework neutrinos deserve dedicated interest. In particular, for the most studied such quantum spacetime, the so-called ``Moyal non-commutative spacetime", it is remarkably found \cite{szabo} that the implications of spacetime quantization for particle propagation end up depending on the standard-model charges carried by the particle and its associated coupling to other particles.
Accurate studies of  (\ref{main}) for neutrinos would be our first opportunity
to tangibly constrain such possibilities for a particle carrying weak-interaction charge.

Testing the applicability of (\ref{main}) to GRB neutrinos is in principle simple.
GRBs last anywhere between a few and $\sim 1000$ seconds
and if $t_{LIV}=0$ the associated neutrinos are, of course, expected
to be detected within approximately the  same time window. Such a  coincidence in arrival time of
GeV photons and sub-Mev photons in some GRBs enabled the Fermi satellite to set limits for $M_{LIV} (photons)\gtrsim M_{P}$
for photon propagation using just this idea \cite{fermiNATURE}.
If instead $t_{LIV}$ is described by (\ref{main}), for sufficiently high energies
and sufficiently high redshifts $t_{LIV}$ would be large,
and the neutrinos would be detected either significantly before or
significantly after the time interval when the low-energy  photons
of the same GRB are observed.
But here resides the challenge\footnote{A somewhat similar description of the challenges for testing (\ref{main}) at IceCube
was given by Gonzalez-Garcia and Halzen a few years ago \cite{halzen2006}. Consistently with what we are here arguing,
they concluded that these challenges would have to be reassessed
once the first data from IceCube could be analyzed.} that most significantly affects the
interpretation of the observations.
If the neutrinos are detected much before or much after the time interval when the GRB is observed
how would we know that they are GRB neutrinos?
There is, as mentioned, a background of other neutrinos (in particular atmospheric neutrinos)
that the telescopes detect. The key discriminator being used in searching for candidate
GRB neutrinos
exploits the
fact that the expected rate of background neutrinos is sufficiently low
that the chances of accidentally catching a background neutrino are negligibly small
when restricting the search to neutrinos from  (roughly) the same direction
of the GRB photons and in a narrow time window around the time of arrival of the signal in photons.
If however $t_{LIV}$ is described by (\ref{main}), also considering that $M_{LIV}$ has, as mentioned, a rather sizable ''theoretical
uncertainty'' { and $E$ has a significant observational error}, the temporal window should be made considerably larger and contending
with background neutrinos may be a severe challenge.
Jacob and Piran \cite{jacobpiran} have addressed this issue for
GRB neutrinos of energies higher than those here of interest.
In that case, they argue, the background noise is sufficiently low that a detection of a neutrino from the direction of a GRB
can be significant\footnote{ Notice however that, as also observed by Jacob and Piran \cite{jacobpiran},
a detection of a single such neutrino is not enough on its own: only a consistent detection of several positionally coinciding and
consistently time shifted  neutrinos from different GRBs would indicate an observation of $t_{LIV}$ as described by (\ref{main}).}
even when there is a sizable mismatch of detection times.

However, even at lower energies one can
efficaciously test (\ref{main})
upon adopting a change of approach such that the selection of GRB-neutrino candidates is based on
rather tight directional criteria (the direction of the neutrinos should be determined to
be rather accurately consistent within the point spread function of the detector with the direction of the GRB potentially associated to it)
while the time-window criteria for the selection of GRB neutrinos should be  relaxed {but in a systematic way
allowing for (\ref{main})}.
If this strategy is adopted we would gain the ability to test both the $t_{LIV}=0$ hypothesis
and the hypothesis that  $t_{LIV}$ be described by (\ref{main}). It should be appreciated
that these two hypotheses would affect the data analyses not only quantitatively but also qualitatively.
If indeed GRBs are sources of TeV neutrinos and   $t_{LIV} \equiv 0$   then at some point we will have quite a few such
directionally-selected GRB-neutrino candidates, and some of them will be established to be definitely
GRB neutrinos because of a level of
time coincidence with the associated GRBs that would allow us to exclude confidently
the possibility of having caught a background neutrino.
On the other hand, if  $t_{LIV}$ is described by (\ref{main}) one should
expect that we might never have
a specific neutrino that can be conclusively associated to a GRB and yet we could deduce that some of the neutrinos
(without knowing which ones) did come from GRBs, just because the distribution of times of detection
of directionally-selected neutrinos would not be just random (as in the case of a sample of pure background
neutrinos): the sample would manifest a higher probability of detecting neutrinos
in a certain energy-dependent { and redshift  dependent } time window, governed by (\ref{main}), systematically advanced or delayed with respect to the gamma-ray trigger of the GRB.

Of course, the accumulation of candidate
GRB neutrinos will provide more or less insight depending on how sharply the energy of the neutrino candidates
is determined experimentally, { on the availability of accurate position and redshift of the GRB } and on the robustness of the inferred value of $t_{LIV}$.
Neutrino energies are determined by IceCube with a $30\%$ uncertainty \cite{halzen}.
{ GRB positions are determined very accuratly if X-ray of optical afterglow is observed and redshift determination
can be achieved if a strong optical afterglow with suitable absorption lines is detected   or if a host galaxy is identified.}
Concerning the robustness of the inferred value of $t_{LIV}$ one should notice that,
according to current models, the emission
of neutrinos should coincide with the GRB trigger in gamma rays
up to a possible advanced emission of  a few tens of seconds
or a possible delay of emission which could go as far as about 100 seconds after the duration of the burst.
It is of course still appropriate to describe $t_{LIV}$ as the difference between the time of detection
of the neutrino and the trigger time of the GRB tentatively associated to it,
but in testing the hypothesis characterized by Eq.(\ref{main}) one should take into
account the uncertainty in the emission time of neutrinos within the GRB event that current models
allow for.
So to each candidate GRB neutrino we should assign
a time offset $t_{LIV} - \Delta t_{GRB}$
where $\Delta t_{GRB}$ reflects the uncertainty GRB modeling attributes to the delay of emission
of neutrinos with respect to the GRB trigger.
In this respect we should stress that in all 3 events that we examine here  the inferred
value of $t_{LIV}$ is significantly larger  than 1000 seconds, so the possibility that the neutrino might have been emitted
a few tens of seconds before the GRB trigger can be neglected.
Instead, the possibility that neutrinos be emitted at any time during the GRB phenomenon
and  up to 100 seconds after the GRB ends can occasionally matter, but only for busts of unusually
 long duration, long enough to make this $\Delta t_{GRB}$ non-negligible with respect to $t_{LIV}$.
Amusingly one of the three GRBs relevant for our analysis is just in this situation:
GRB090417B had an unusually long duration of some 2300 seconds,
so for the $t_{LIV}$ of the neutrino tentatively associated to GRB090417B
we shall allow for a $\Delta t_{GRB}$ of 2400 seconds.
While such long GRBs are rare,  similar cases may present themselves again as more GRB-neutrino candidates are accumulated.

Once data are collected following these criteria a first level of assessment in relation
to the content of Eq.(\ref{main}) can be given in the spirit  illustrated in
Fig.~1. With a large number of directionally-selected GRB-neutrino candidates
one could conclusively test (\ref{main}), even without any redshift information about the relevant GRBs
and even if each individual event had a nonnegligible chance of being a background event.
Fig.1 conveys this message by taking as illustrative example the case  $s_\pm =1$ and  $M_{LIV}=0.1M_P$.
The shaded area of Fig.~1 shows how these illustrative hypotheses would affect
 the prediction from (\ref{main}) of the correlation between values of the energy
 of the candidate GRB neutrinos and of the inferred value of $t_{LIV}$, given in terms of
 the time-of-detection difference with respect to the trigger of the
 lower-energy photon signal of the relevant GRB.
 The thickness of the shaded area in Fig.~1 reflects the simplifying assumption
that the redshifts of the candidate GRB sources of the neutrinos all take value between 0.2 and 8.
(This broad redshift range encompasses more than 95\% of the long GRBs with known redshifts; the darker part of the
shaded region of Fig.~1 would apply to GRBs in the narrower redshift range from 0.9 to 3, which contains
about 50\% of values found for long GRBs with known redshifts).

\begin{figure}[htbp!]
\includegraphics[width=0.95 \columnwidth]{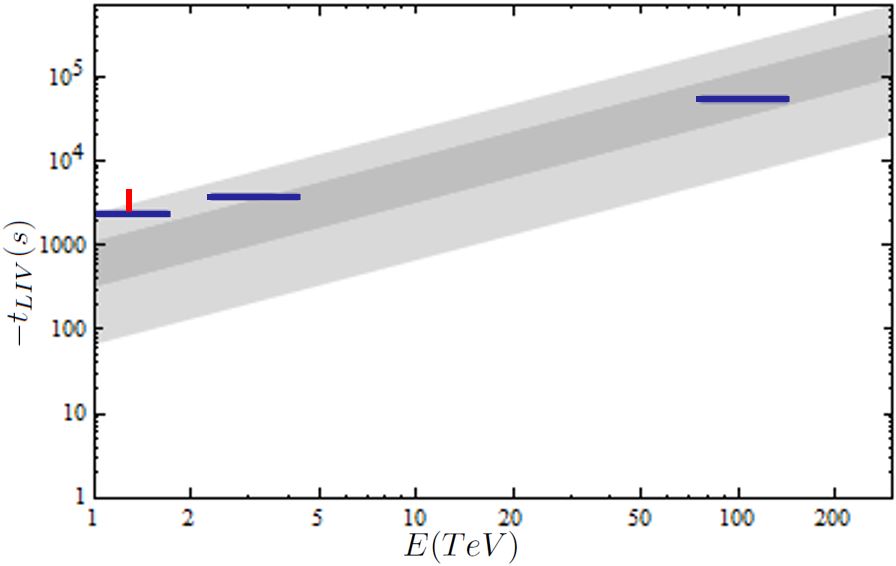}
\caption{The conjetured quantum-spacetime effects of Eq.(\ref{main})
can be cast in the form of a prediction for correlations between energy of GRB neutrinos
and the value of $t_{LIV}$ that can be inferred from
their detection time (as compared to the trigger in gamma rays of the detection of the associated GRB).
According to (\ref{main}), assuming as illustrative example the case with  $s_\pm =1$ and $M_{LIV}=0.1 M_P$,
such correlations should all fall within the shaded region for all GRB neutrinos,
as long as the redshift of the GRB source is between 0.2 and 8 (and in particular within the darker part of the
shaded region for GRBs with redshift between 0.9 and 3).
The 3 IceCube events  are shown  with the horizontal segments reflecting
 a $30\%$ uncertainty in their energy. The vertical red segment reflects our estimate of $\Delta t_{GRB}$, which
is appreciable for the lowest-energy event, tentatively associated to an unusually long burst.}
\end{figure}

 While at first glance in Fig.~1, all the 3 GRB-neutrino candidates
fit within the shaded region, this figure essentially factors out all information
on redshifts of the candidate sources.
The assessment of these 3 candidate GRB neutrinos
can be made more precise
by imposing the consistency condition that a group of genuine GRB-neutrino events governed
by (\ref{main}) should all be consistent with the same value of $M_{LIV}$ when taking into account
the (however partial) information available on their redshift.

{As an illustration of a Lorentz invariance violation inspired analysis of the data we examine here the scenario in which the three IceCube events
are GRB neutrinos and the arrival times are determined by  (\ref{main}).}  Fig.~2a depicts the allowed range of redshifts and $M_{LIV}$ for each of the neutrino events, assuming
$t_{LIV}$   given by  (\ref{main}).
Fig.~2b (respectively 2c)  describes the probability that an observed
long (respectively short) burst is at a certain redshift\footnote{A more precise estimate can be obtained using additional information concerning a burst, such as its fluence.}, on the basis of the
observed redshift distribution for the long GRBs  \cite{wandeLONG} and for the short GRBs \cite{cowaSHORT}.

\begin{figure}[htbp!]
\includegraphics[width=0.9 \columnwidth]{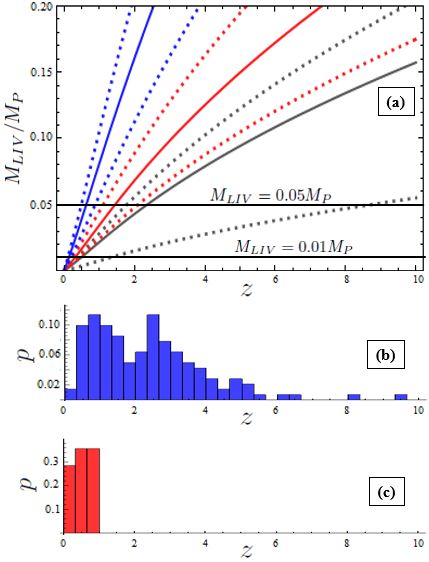}
\caption{For given
   $t_{LIV}$  and neutrino energy  (\ref{main}) is consistent with
a combination of values of redshift and $M_{LIV}$.
This is shown with the continuous lines in panel (a): blue is for 109 TeV and $t_{LIV}$ of 14 hours,
red is for 3.3 TeV and $t_{LIV}=3594s$, and gray is for 1.3 TeV and $t_{LIV}=2249s$.
Dashed lines delimit the range of uncertainty due to the uncertainty in the energy determinations
and $\Delta t_{GRB}$ (which is appreciable only for the lowest dotted gray line).
Panel (b) (respectively (c))  describes the probability that an observed
long (respectively short) burst has a certain redshift. For $M_{LIV} \lesssim 0.05M_{P}$ it is plausible to interpret both the
 $109 TeV$ and the $3.3TeV$ events as GRB neutrinos governed by (\ref{main}). For higher values of $M_{LIV}$ taking the $3.3TeV$ event as
  a GRB neutrino requires assuming a redshift for short burst GRB090219 that is unlikely on the basis of
  panel (c).
  Interpreting the $1.3TeV$ event as a GRB neutrino requires
  values of $M_{LIV}$ no greater than $\sim 0.01M_{P}$, since in panel (a) this allows the 1.3TeV event to be associated to
  a source at redshift of $0.35$ (as established for GRB090417B).
  For $M_{LIV} \sim 0.01M_{P}$ one can also interpret the $3.3TeV$ event as a GRB neutrino
  governed by (\ref{main}), since then the inferred value of redshift for short burst GRB090219 is consistent with panel (c).
However at values of $M_{LIV}$ as low as $\sim 0.01M_{P}$ the interpretation of the 109 TeV in association to the long burst GRB091230A
   is  disfavored by the probability distribution shown in panel (b).}
\end{figure}

A horizontal line in Fig.2a corresponds to a given value of $M_{LIV}$ and for illustrative purposes two
such lines are shown.
Each line implies a range of redshift values (corresponding to the one-standard-deviation energy range of each one of the neutrinos)
for each one of the bursts. These ranges should be compared with the expected probabilities that the bursts have a given redshift as seen by Figs.~2b and 2c. Here one should distinguish between GRB090417B and GRB091230A that are long GRBs with an observed redshift distribution  extending from 0.1 to 9.4 and peaking
in the region $0.5<z<3$, and GRB090219, a short GRB whose expected redshift is between 0.1 and 1.

{ It is remarkable that even with the limited data available concerning these three events some conclusions can be drawn from this analysis.}
One sees from Fig.~2  that the hypothesis that all 3 neutrinos are GRB neutrinos  is compatible with (\ref{main}) only if we assume a corresponding hierarchy for the redshifts of the candidate sources and if we allow some (in cases two of the events) to be
attributed very unlikely redshift values.
This hypothesis would lead to the assumption that   GRB091230A  was closer than both GRB090417B  and GRB090219, the farthest being
GRB090417B. However,
GRB090417B was a very long optically-dark burst with a rather robust association with a SDSS galaxy at
  redshift  $z \simeq 0.35$ \cite{ultralong}.
It would seem rather implausible
that for GRB091230A, a long GRB whose redshift was not determined, had a redshift smaller than 0.35 (less than 2\% of the observed long GRBs have
$z<0.35$). Therefore the possibility that all three events are associated with GRBs is unlikely.
On the other hand for GRB090219, which was a short burst, it is rather plausible to assume that $z<0.35$
(about 30\% of short GRBs have $z<0.35$). It is therefore reasonable to contemplate the hypothesis that
both the 1.3TeV event and the 3.3TeV event be GRB neutrinos. This would require $M_{LIV} \lesssim 0.01M_P$.
Another reasonable possibility is that the 1.3TeV event was background, but
both the 109 TeV event and the 3.3TeV event were GRB neutrinos.
Making the reasonable assumption
that the long burst GRB091230A was at redshift $0.4 < z < 5.5$
(see Fig.~2b) one gets a range of compatible values of $M_{LIV}$:  $0.02M_P< M_{LIV} < 0.5M_P$.
While  the 3.3TeV candidate from the short burst GRB090219 implies  $M_{LIV} \lesssim 0.05M_P$.
Combined together we find that for $0.02 M_P \lesssim M_{LIV} \lesssim 0.05 M_P$ both the  109 TeV
event and the 3.3TeV event could be tentatively considered as GRB neutrinos.

In summary we conclude that
at most 2 of the 3 GRB-neutrino candidates could possibly be GRB neutrinos governed by (\ref{main})
{ and if so
this points towards a rather low value for $M_{LIV}$for neutrinos.}
We stress that the hypothesis that one or two of the candidate events might
be GRB neutrinos governed by (\ref{main})
should only be viewed within the realm of plausibility,
the most likely interpretation of the data being of course that all 3 candidates
are insignificant background events.
Nonetheless, going back to a key point we made earlier, it is interesting that
we have here a quantum-spacetime model with independent reasons of interest
within which one gets a rather plausible interpretation of presently
available IceCube data as including perhaps as many as 2 GRB neutrinos.

While unlocking the secrets of the mechanisms producing  neutrinos at GRB sources is of
very high intrinsic interest,  in assessing the motivation for future efforts along this line of analysis
one should also take into account that the possible confirmation of the Lorentz invariance violation
contemplated  here would have a gigantic impact on  fundamental physics.
One would not only have the first much-sought evidence of a quantum property of spacetime
  (and evidence imposing at least an adaptation of Einstein's special relativity to that quantum spacetime context),
  but one would also have a very intelligible hint concerning the correct description of quantum spacetime.
  This is particularly true since tests of the applicability of Eq.(\ref{main})
  to photons
from  GRB090510
observed by the Fermi gamma-ray telescope  
lead to a bound  for photons of
$M_{LIV}(photons) > 1.2 M_P$ \cite{fermiNATURE}. If indeed for neutrinos future data
ended up providing evidence for $0.01 M_P \lesssim M_{LIV} \lesssim 0.1 M_P$ this would immediately direct us
toward the few models, mentioned earlier, in which  the laws
of particle propagation in quantum spacetime depends on either the spin of the particle or
its standard-model charges.

We stress that  to further investigate this scenario
some measures should be adopted at neutrino telescopes such as IceCube.
Specifically one should consider a joint analysis of candidate events that are localized within
the direction of GRBs but shifted in time, searching for a possible common interpretation
with a single $M_{LIV}$ value, in the way presented here.
One can view this new suggested analysis in Baysian like approach.  Lacking an exact model for GRB neutrino emission and given that other sources apart from the source of the prompt emission could produce high energy GRB neutrinos, standard analyses assume a prior according to which GRB neutrinos should emerge more or less uniformly within about a 100 sec after the trigger. Inclusion of possible neutrino emission during the jet propagation within a Collapsar should shift this prior back by  $ \sim 20-30 $ seconds prior to the trigger. The new analysis that
considers $t_{LIV}(E,z,M_{LIV})$ should adopt
a new prior in which this original duration is modified according to (\ref{main}). For a given data set that will include neutrino candidates from GRBs with and without redshift one should then estimate the probability of association of neutrinos to GRBs according to the standard old prior but also using  this new one. In principle a significant fit with this new prior could
point towards Lorentz  invariance violation and enable us to estimate $M_{LIV}$.

\bigskip

We thank Francis Halzen, for several helpful discussions, and Fabrizio Fiore, for valuable comments on an early
 version of the manuscript. This research was partially supported by the John Templeton Foundation (GAC) and by an ERC advanced grant (TP). TP thanks the IUSS - Ferrara for hospitality while this research was done.

\end{document}